# Magnetic field tuning of coplanar waveguide resonators

J E Healey[1], T Lindström[1,2], M S Colclough[1], C M Muirhead[1] and A Ya Tzalenchuk[2]

[1] University of Birmingham, Edgbaston, Birmingham B15 2TT, UK

[2] National Physical Laboratory, Hampton Road, Teddington, TW11 0LW, UK

We describe measurements on microwave coplanar resonators designed for quantum bit experiments. Resonators have been patterned onto sapphire and silicon substrates, and quality factors in excess of a million have been observed. The resonant frequency shows a high sensitivity to magnetic field applied perpendicular to the plane of the film, with a quadratic dependence for the fundamental, second and third harmonics. Frequency shift of hundreds of linewidths can be obtained.

Thin film resonator structures are used extensively for microwave applications. Those based on superconductors have the advantage of very sharply defined resonances because of the very low conductor losses, and have found application in filters for telecommunications[1,2] and astrophysical detectors[3,4], among which kinetic inductance detectors[5,6] are particularly relevant to the present work.

Recently, superconducting coplanar resonators (CPRs) have also been used as tuned cavities to interrogate and manipulate the state of quantum bits (qubits)[7]. When the resonant frequency of a cavity and qubit are close, the quantum state of the system becomes a hybrid of the cavity and qubit states enabling fundamental studies of quantum electrodynamics, applications to quantum metrology and quantum information science. The reported work in the cavity-qubit system has involved charge[8], phase[9], or hybrid qubits[10,11]. The ability to adjust the CPR resonant frequency ($f_0$) while maintaining the qubit at the optimal operating point would be of considerable value. Schemes such as the one for generating single photons on demand using the flux-qubit/CPR system described in a recent paper[12] would be greatly simplified and improved if the resonator could be frequency-tuned independent of the qubit. This would also allow the possibility of bringing multiple resonator structures[13] in and out of resonance with each other.

The important feature of any such control system is that it should not itself introduce substantial excess losses. There have been several reports of techniques to adjust the resonant frequency of CPRs using the non-linear inductance of current controlled Josephson junctions[14,15] or field-controlled SQUIDs[16,17]. There is also extensive reported work on the use of ferroelectric layers[18] and, more recently, ferroelectric-ferromagnetic hybrids to adjust the centre frequency of superconducting resonators[19]. Even though these approaches provide wideband tunability, they appear to introduce significant additional losses into the system, even when the control is not activated, and therefore have limited use in applications requiring very high quality factors (Q).

In this paper we describe low temperature experiments with high-Q CPRs in very weak magnetic fields. Small non-linear effects in the superconductor produce large changes in the centre frequency compared to its linewidth, without any additional losses.

The CPRs were all fabricated from 200 nm thick niobium films sputtered onto 0.5 mm thick R-plane sapphire and onto thermally oxidised high resistivity silicon substrates. They were patterned with conventional photolithography and argon ion beam milling. The structure is shown in Figure 1. The frequency response of the resonators was measured with a vector network analyser. At low temperature we obtain the CPR centre frequency of around 5.7 GHz for sapphire and 6.0 GHz for silicon substrates.

Two measurement systems were used. Measurements from 20 mK to 2 K were made in a dilution refrigerator fitted with a cold InP HEMT[20] preamplifier having a bandwidth of 4-8 GHz. This system allows observation of only the fundamental resonance, but has high sensitivity and temperature stability. Magnetic field could only be applied at an angle $\phi$ of about 10° to the plane of the substrate.

Measurements from 1.2 to 4.2 K were made in a glass cryostat. This simple system has the advantage that in additional to the fundamental, the second and third harmonic can also be measured and small magnetic fields can easily be applied in any direction via Helmholtz coils external to the cryostat.

Temperature dependence of the quality factor and the resonance frequency of one of the resonators fabricated on sapphire is shown in Figure 2. Loaded Q up to $9 \times 10^5$ independent of the excitation power has been measured on these resonators, depending on the coupling dimensions W & G in the central line of coplanar waveguides. We were able to modify the loading controllably by trimming the width W of the gaps using focused ion beam milling. By trimming the feed line side of the gap this could be achieved with no noticeable change in the resonance frequency.

The high-temperature behaviour of both Q and $f_0$ are well understood in terms of the change in the complex microwave surface impedance associated with thermally excited quasiparticles. As the temperature is decreased, the number of thermally excited quasiparticles also decreases leading to a lower loss (higher Q) and smaller kinetic inductance (higher $f_0$). The behaviour below about 1.5 K is characterized by a slower increase in Q but a *decrease* in $f_0$. Such behaviour has previously been observed[21], but has not found a generally accepted explanation. There is agreement, however, that losses at low temperatures are limited by a dissipation mechanism unrelated to superconductivity and Ref 21 has found good agreement between two-level fluctuator theory and the temperature dependence of both $f_0$ and Q for a CPR on Si. We have seen similar behaviour in resonators fabricated on both sapphire and silicon substrates. We note that the shape of the temperature dependences is reminiscent of the effects of paramagnetic impurities contributing a Curie term proportional to 1/T to the magnetic susceptibility of the substrate. This contribution is exploited in very high-Q whispering gallery mode sapphire resonators for compensation of the thermal expansion/contraction of the lattice[22].

The behaviour in the samples fabricated on silicon is further complicated by a strong dependence of Q on the power of microwave radiation. We have measured loaded Qs up to $1.1 \times 10^6$ at the lowest temperatures using microwave excitation power strong enough to saturate fluctuators in the substrate. Reduction of the excitation power by only 6 dB at a few hundred mK resulted in an order of magnitude reduction in Q. Relatively insignificant power dependence of Q was observed in the samples fabricated on sapphire, which reflects the much lower concentration of defects and impurities in this substrate. We will discuss these observations in more detail elsewhere. We turn now to the dependence of resonant frequency on weak applied magnetic fields, which is the main result of this paper.

We have measured the field response of 20 resonators, of which 6 were on silicon and 14 on sapphire, and observed some spread in the *magnitude* of the field response, but qualitatively all datasets showed the same behaviour. There is no evidence that the magnitude of the field response is related to the substrate. What is different between the samples is the patterning process to which we would expect the coplanar structure to be particularly sensitive because the microwave currents are confined to within about a penetration depth of the film edges.

Figure 3.a shows the dependence of the resonance frequency of a resonator on the magnetic field applied at a small fixed angle ($\phi \sim 10°$) to the plane for a CPR on a silicon substrate. The dependence is almost exactly quadratic with a slight offset along the field axis due to flux trapped in the film. This offset was circumstantial and is absent in other datasets. Sensitivity $df_0/dH \, \alpha \, H$ was independent of temperature, field sweep direction, or microwave power, as shown in Figure 3.b. Crucially, Q was insensitive to the applied magnetic field, Figure 3 c.

The Q in this temperature range is limited by factors extrinsic to the superconductor so the absence of significant field dependence is to be expected. Identification of these limiting factors is clearly important for the reduction of decoherence in quantum systems coupled to the resonator.

In Figure 4, we show the change in resonant frequency as a function of magnetic field applied *perpendicular* to the plane of the film for the fundamental, second and third harmonics of a sample on sapphire at 1.3 K, having a Q of $3.10^5$. The resonant frequency could be tuned reproducibly and reversibly by as much as 2 MHz : this is ~ 200 bandwidths for this resonator, with an applied field of only 0.2 mT. Comparable data has been obtained at 50 mK in resonators having Q close to $10^6$. In the inset we show the frequency shift for the fundamental mode at fixed applied field as the field direction is rotated about the long direction of the CPR. There is a closely sinusoidal dependence on angle φ, with no suppression within experimental error when the field is applied parallel to the plane of the film for θ=0° or θ=90°. This angular dependence is easily understood in terms of flux-focusing by the large demagnetising factor for the perpendicular field component. We see also that the frequency shift is larger for the higher harmonics in the ratio $f_3:f_2:f_0$=2.8:1.8:1. This is slightly different to the 3:2:1 ratio predicted by our model, but we note that the flux focusing can be expected to be slightly non-uniform near the ends of the CPR, and the different resonant modes will be sensitive to different parts of the sample.

An $H^2$ dependence can be explained by non-linear London equations[23,24], which lead to an expression for the frequency $f(T,H) = f(T,0)\left(1 - (L_K(T,0)/L_T)\beta(T)H^2/H_C^2\right)$, where $L_K(T,0)$ is the zero-field kinetic inductance, $L_T$ is the total inductance, β(T) is a scaling factor, $\beta(T)H^2$ is small compared to $H_C^2$ and $L_K$ is small compared with $L_T$. Using the normalized frequency dependence in Figure 2 and data extending to 4.2 K, we find an approximate ratio for $L_K/L_T$ of 1.5%. Small limit behaviour is well satisfied for the data shown in Figure 3. A quadratic dependence for the penetration depth has been reported in both conventional[25,26,27] and high temperature[28] superconductors.

For the sample in Figure 4. we find that the quadratic behaviour extends up to an applied field of 0.25 mT, above which the Q becomes seriously degraded and the resonant frequency is depressed even after return to zero applied field. A small field of opposite sign will then slightly increase the resonant frequency. Flux clearly moves in the film. If we identify an applied field of 0.25 mT for this sample with the lower critical field for niobium $B_{c1}$~0.1 T, this implies a flux focusing factor ~ 400, which is not at all unreasonable considering the very high demagnetizing factor in the field perpendicular direction. This implies that the quadratic behaviour is maintained even though H is no longer small compared with $H_C$.

The theory of Ref. 24 predicts that for conventional superconductors $\beta(t)$ goes to zero at T=0, reflecting the BCS decrease in quasiparticle density. In contrast we find for our samples only a 5% reduction of $\delta f$ =f(T,0.23 mT) –f(T,0) from 4.2 down to 1.3 K, and no further change down to 20 mK. By using the gap as a variable parameter, Tinkham[29] and Bardeen[30] were able to reproduce the $H^2$ dependence at finite values of the gap (and by implication finite $\beta(T)$), even as T$\rightarrow$0. The experimental situation is complex with work on metallic superconductors showing a $\delta f$ which, at low temperatures, falls rapidly[31], flattens out[23], and increases or changes sign[32,26]. There is a clear need for further investigation of our niobium thin films.

Finally we have investigated the effect of passing current through a patterned control line of only 0.88% the length of a resonator. A *fixed* perpendicular magnetic field of 0.122 mT from a solenoid was used to provide a nonzero $df_0/dH$. At 1.3 K we obtain a current sensitivity 344 Hz/mA up to a maximum frequency shift of ±85 kHz, limited by vortex generation – over 3 linewidths for this resonator.

In summary, we can suppress the fundamental resonant frequency of niobium CPRs reversibly by ~ 200 linewidths with a weak perpendicular magnetic field, by exploiting the very large flux focusing factor in this geometry. We have also demonstrated suppression greater than a linewidth by current in a very small control line. In contrast to methods based on ferroelectric layers or on non-linear Josephson elements, suppression can be achieved with no observable reduction of a Q in excess of $10^6$. Optimized control lines may enable fast and efficient frequency tuning. This augers well for applications to qubit-CPR systems, kinetic inductance detectors and a broad range of microwave filter applications.

We are grateful to Phil Meeson, Gregoire Ithier, Giovanna Tancredi, Michael Lancaster, Ed Tarte, David Cox, Carol Webster, Olga Kazakova, John Gallop and Mark Oxborrow for many illuminating discussions and practical help.

**Figure 1.** Sketch of a coplanar resonator. Angles indicate the in-plane, θ, and out-of-plane, ϕ, orientation of the magnetic field with respect to the substrate. Inset: Photograph of the coplanar resonator, comprising a centre conductor of length 11 mm and width W=10 μm separated from ground planes by an S=5 μm gap. The centre conductor is capacitively coupled to the feed lines by gap G of 4 – 8 μm at each end. The meander line is used to confine the whole structure within the 10 mm long substrate.

**Figure 2.** Dependence of resonant frequency and quality factor Q on temperature for an overcoupled CPR fabricated on a sapphire substrate.

**Figure 3.** Centre frequency shift with respect to the centre frequency at B=0, T=25 mK, and –21 dBm power measured at the output of the VNA (about –80 dBm on the sample), $f_0$-6.0376 GHz (a), quality factor (b) and sensitivity to magnetic field (c) of CPR as a function of magnetic field applied at a small angle ϕ~10º to the plane of a Si substrate at various temperatures, two excitation powers and two directions of the field sweep.

**Figure 4.** Change in resonant frequency with perpendicular magnetic field for the fundamental, first and second harmonics. The data is highly reproducible with changing magnetic field. The inset shows the dependence of fundamental frequency shift on the angle ϕ at 0.2mT.

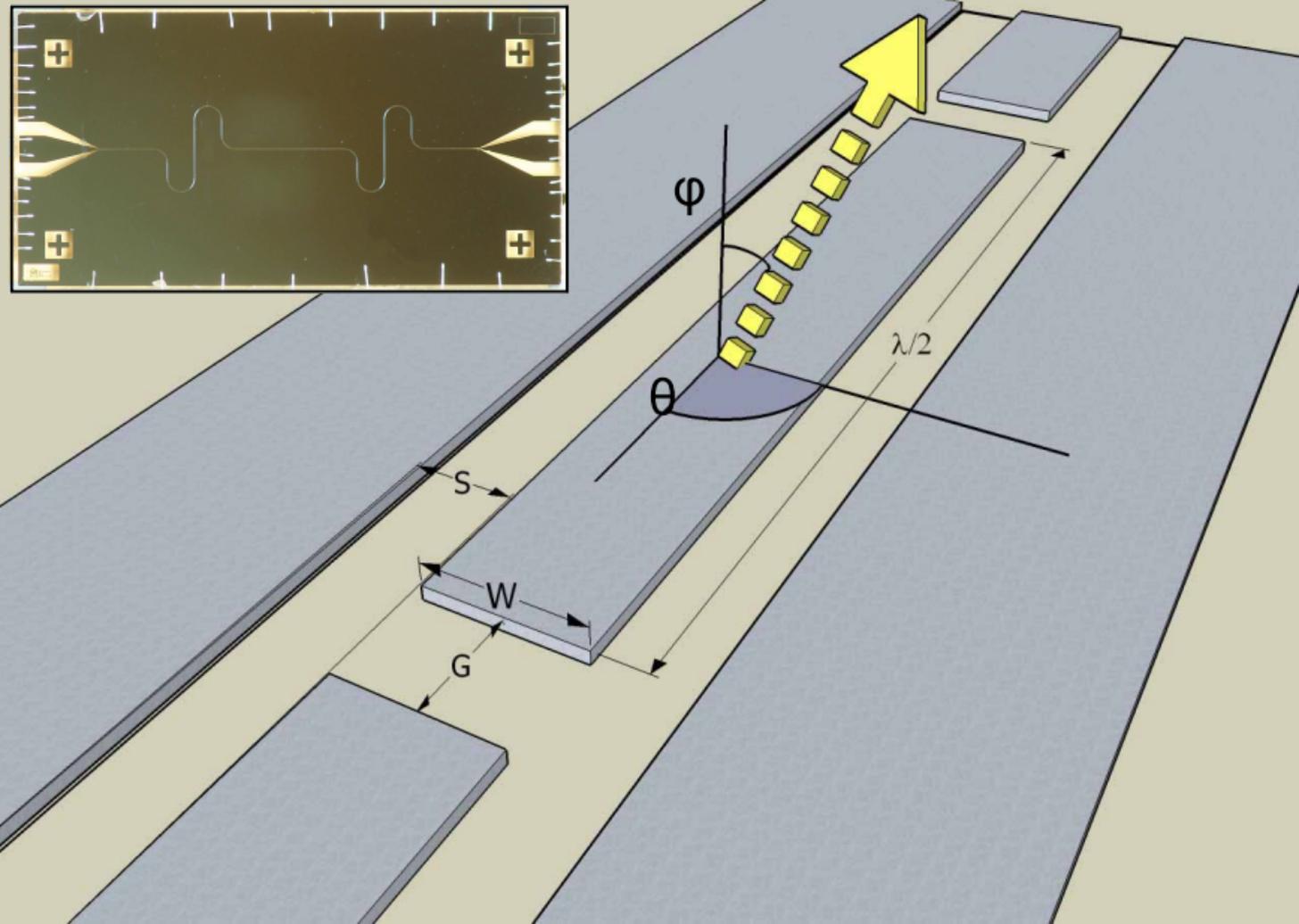

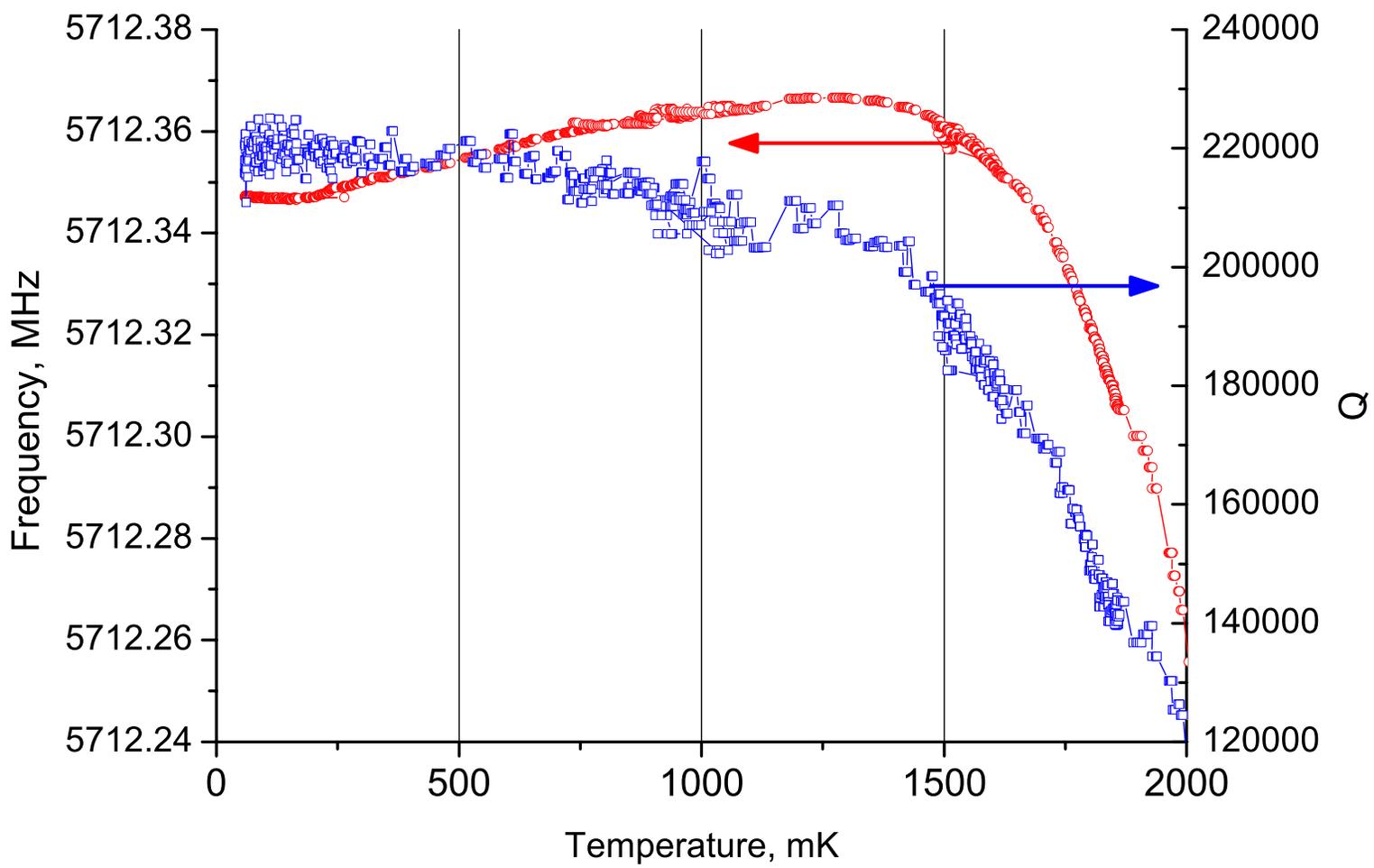

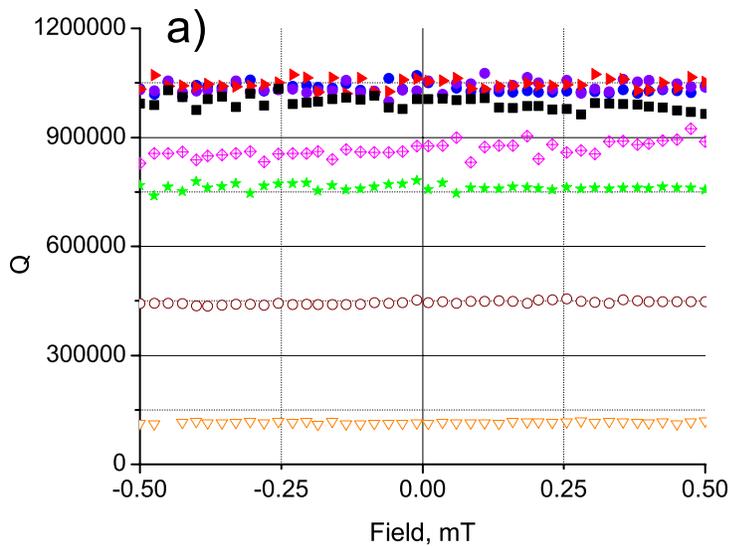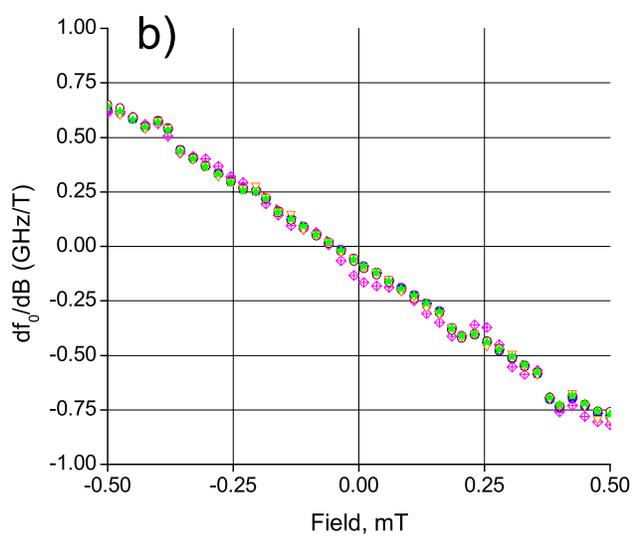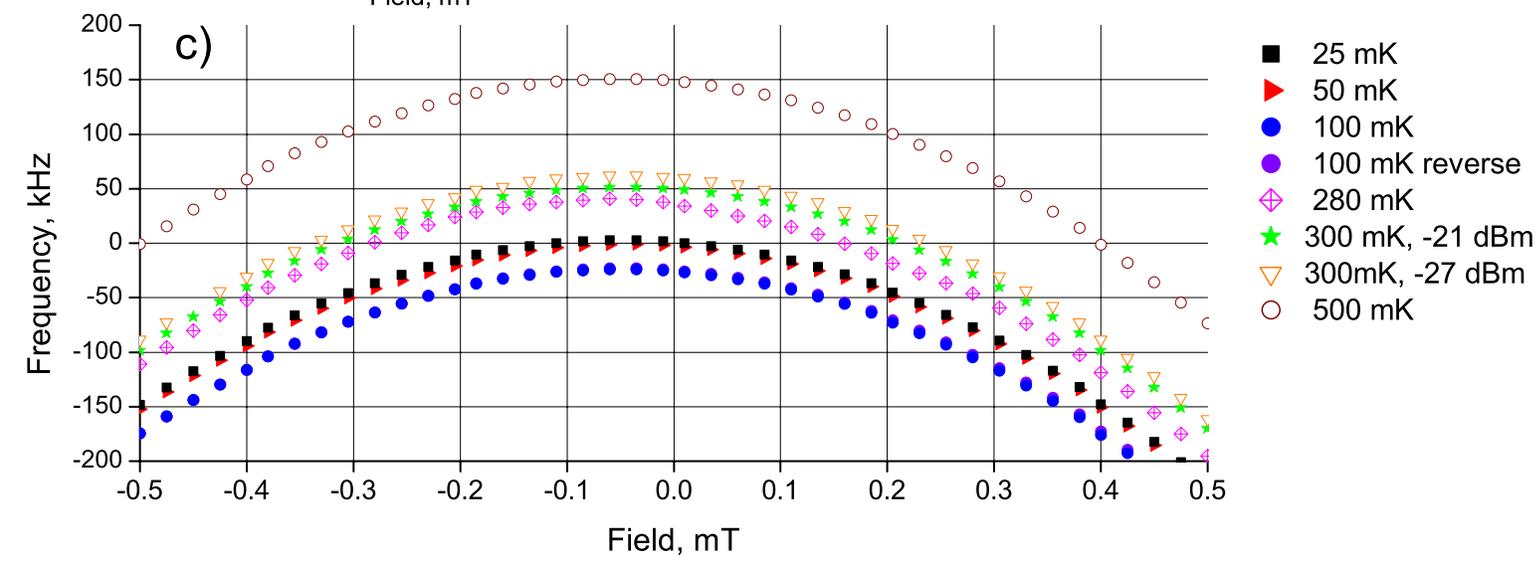

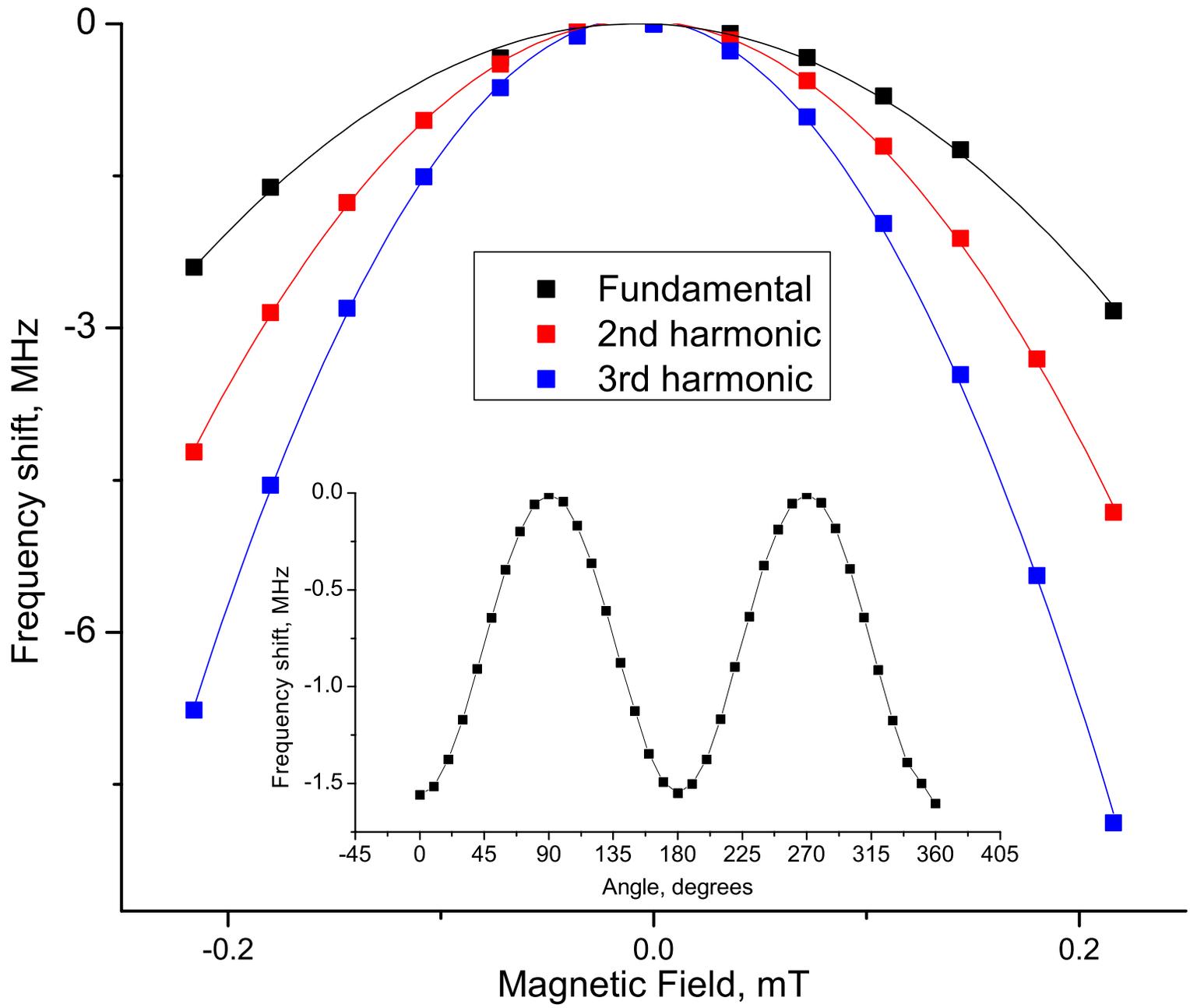